\documentclass[12pt]{article}
\usepackage{amsmath}
\usepackage{latexsym}
\usepackage{amsfonts}
\usepackage[normalem]{ulem}
\usepackage{soul}
\usepackage{array}
\usepackage{amssymb}
\usepackage{extarrows}
\usepackage{graphicx}
\usepackage[backend=biber,
maxbibnames=99,
giveninits=true,
style=numeric,
sorting=none,
isbn=false,
doi=false,
url=false
]{biblatex}
\AtEveryBibitem{\clearlist{language}}
\usepackage{wrapfig}
\usepackage{wasysym}
\usepackage{enumitem}
\usepackage{adjustbox}
\usepackage{ragged2e}
\usepackage[svgnames,table]{xcolor}
\usepackage{tikz}
\usepackage{longtable}
\usepackage{changepage}
\usepackage{setspace}
\usepackage{hhline}
\usepackage{multicol}
\usepackage{tabto}
\usepackage{float}
\usepackage{multirow}
\usepackage{makecell}
\usepackage{fancyhdr}
\usepackage{indentfirst}
\usepackage[font=small,labelfont=bf,justification=justified]{caption}
\usepackage[toc,page]{appendix}
\usepackage[hidelinks]{hyperref}
\usetikzlibrary{shapes.symbols,shapes.geometric,shadows,arrows.meta}
\tikzset{>={Latex[width=1.5mm,length=2mm]}}
\usepackage{flowchart}\usepackage[paperheight=11.0in,paperwidth=8.5in,left=1.0in,right=1.0in,top=1.0in,bottom=1.0in,headheight=1in]{geometry}
\usepackage[utf8]{inputenc}
\usepackage[T1]{fontenc}
\TabPositions{0.5in,1.0in,1.5in,2.0in,2.5in,3.0in,3.5in,4.0in,4.5in,5.0in,5.5in,6.0in,}
\usepackage[version=3]{mhchem} 
\usepackage{chemformula}
\usepackage{threeparttable}
\usepackage{svg}
\usepackage[justification=centering]{subfig}
\usepackage{outlines}
\usepackage{setspace}
\usepackage{breqn}
\usepackage{xcolor}
\usepackage{listings}
\AtEveryBibitem{%
  \clearfield{note}%
}

\usepackage[labelsep=period]{caption}
\usepackage{graphicx}

\renewcommand{\_}{\kern-1.5pt\textunderscore\kern-1.5pt}

\setlistdepth{9}
\renewlist{enumerate}{enumerate}{9}
        \setlist[enumerate,1]{label=\arabic*)}
        \setlist[enumerate,2]{label=\alph*)}
        \setlist[enumerate,3]{label=(\roman*)}
        \setlist[enumerate,4]{label=(\arabic*)}
        \setlist[enumerate,5]{label=(\Alph*)}
        \setlist[enumerate,6]{label=(\Roman*)}
        \setlist[enumerate,7]{label=\arabic*}
        \setlist[enumerate,8]{label=\alph*}
        \setlist[enumerate,9]{label=\roman*}

\renewlist{itemize}{itemize}{9}
        \setlist[itemize]{label=$\cdot$}
        \setlist[itemize,1]{label=\textbullet}
        \setlist[itemize,2]{label=$\circ$}
        \setlist[itemize,3]{label=$\ast$}
        \setlist[itemize,4]{label=$\dagger$}
        \setlist[itemize,5]{label=$\triangleright$}
        \setlist[itemize,6]{label=$\bigstar$}
        \setlist[itemize,7]{label=$\blacklozenge$}
        \setlist[itemize,8]{label=$\prime$}

\begin{document}

\captionsetup[figure]{labelfont={bf},labelformat={default},labelsep=period,name={FIG.}}

\section*{Combining ozone and UV-light for energy efficient removal of \ch{SO_x} in low-temperature flue gas}
\addcontentsline{toc}{section}{Combining ozone and UV-light for energy efficient removal of \ch{SO_x} in low-temperature flue gas}

\noindent
Reference number: 19-422 \\ \\
Marc Rovira, Klas Engvall, and Christophe Duwig \\
\textit{ Department of Chemical Engineering, KTH Royal Institute of Technology, SE-10044 Stockholm, Sweden} \\

\noindent
\textbf{ABSTRACT}

The potential of a combined ozone \ch{O3} ultraviolet (UV) light reactor for gas-phase oxidation of flue gas pollutants has been evaluated in this work. For this, numerical simulations of a continuously stirred tank reactor (CSTR) have been performed for the analysis of sulfur dioxide \ch{SO2} removal. Chemical kinetics have been coupled with modeling of the radiation intensity distribution produced by a UV lamp with five different radiation models. A grid convergence study and an optimization of the number of discrete point light sources employed for some of the radiation models was carried out. The effect of residence time and reactor size on the removal of \ch{SO2} have also been analyzed. Overall, results for the abatement of \ch{SO2} in an \ch{O3}-UV reactor under optimized conditions suggest that this approach is suitable for sustainable air cleaning applications.

\subsection*{1. Introduction}
\addcontentsline{toc}{subsection}{1. Introduction}

There is currently a wide consensus that poor air quality leads to adverse heals effects in humans. The World Health Organization estimates that 90\% of the world population lives in areas that do not meet their air quality guidelines and attributes more than 3 million premature deaths each year to outdoor airborne emissions \cite{world_health_organization_ambient_2016}. One of the six priority pollutants considered by international air standards is the family of gaseous sulfur oxides (\ch{SO_x}) of which sulfur dioxide (\ch{SO2}) can be found in much greater concentration \cite{curtis_adverse_2006}. Health effects of \ch{SO2} exposure are more pronounced in susceptible populations such as children and the elderly. For children, increased asthma \cite{lee_air_2002, penardmorand_long-term_2005}, rhinitis \cite{penardmorand_long-term_2005}, pneumonia and acute bronchitis \cite{barnett_air_2005} hospitalizations are linked to higher \ch{SO2} levels. The environment is also affected by high \ch{SO2} emissions mainly through acid rain. 

Since the USA announced the first version of the Clean Air Act in 1963, there has been a global effort to reduce \ch{SO2} pollution from human sources as anthropogenic emissions account for close to 99\% of total emissions \cite{wang_123_2018}. The main industrial sources include vehicle and refinery emissions as well as emissions from the burning of coal \cite{godish_air_2014}. Furthermore, the presence of \ch{SO2} in exhaust gasses is proportional to the concentration of sulfur in the fuel employed. Although some industries such as the automotive have been quick to tackle \ch{SO2} emissions, others like the shipping industry are still large pollutant sources. In 2014, the marine industry was found to be responsible 13\% of global \ch{SO2} emissions \cite{smith_third_2014}. Although many vessels have embraced low sulfur emissions, upcoming legislation is bound to be even more strict. The regulations in the MARPOL Annex VI have been adopted by the International Maritime Organisation (IMO) and define the extent of admissible air pollution impact of ships \cite{kattner_monitoring_2015}. Taking effect in 2020, the revised MARPOL Annex VI reduces the  ``Global Sulfur Cap'' from 3.5\% to 0.50\% and 0.1\% in sulfur Emission Control Areas \cite{animah_compliance_2018}. Hence, the shipping industry will require drastic measures to meet these targets. A similar trend is seen in industrial flue gas emissions \cite{gholami_technologies_2020, sun_abatement_2016}.

Apart from employing low-sulfur fuels when possible, most industries remove \ch{SO2} by employing gas cleaning systems. In these solutions \ch{SO2} is treated along with \ch{NO_x}. However, most implementations do so individually. For flue gas, independent denitration (i.e. selective catalytic reduction (SCR) with ammonia) and desulfurization (i.e. wet flue gas desulfurization WFGD) systems result in costly technologies that are expensive to operate and install \cite{zhao_simultaneous_2010, liu_photochemical_2011}. Hence, multi-pollutant removal solutions are quickly becoming a prominent research topic. One such technology, which uses oxidation by ozone (\ch{O3}) to treat exhaust gases, was recently review by Lin \textit{et al.} \cite{lin_flue_2020}. Although this method achieves high denitrification efficiency, its main limitation of is that \ch{O3} selectively oxidises \ch{NO} \cite{sun_simultaneous_2011}. Hence, the direct oxidation of \ch{SO2} by ozone is not significant. Current solutions employ pre-ozonation to increase the oxidation state of \ch{NO} and thus its solubility to improve removal before a wet scrubber \cite{lin_flue_2020}. Although this process can achieve high removal efficiencies, the treatment of the residual liquid employed in the scrubbing process increases operational costs and limits its applicability. Hence, although oxidative removal of \ch{NO} by \ch{O3} is feasible, other alternatives for the simultaneous oxidation of \ch{SO2} are required.

Advanced oxidation processes (AOPs) that use hydroxyl radicals (\ch{^.OH})  for water purification and disinfection have been employed for wastewater treatment for over three decades \cite{deng_advanced_2015, chong_recent_2010}. Recently, this knowledge has been applied to the gas-phase removal of exhaust gas pollutants with notable success \cite{xie_simultaneous_2019, liu_simultaneous_2018, hao_establishment_2016}. Here, direct exposure of the primary oxidant, usually hydrogen peroxide (\ch{H2O2}), to ultraviolet (UV) light at a particular wavelength creates \ch{^.OH}. This radical is then able to effectively simultaneously oxidize and remove multiple pollutants. Benefits of gas-phase AOPs include low operational costs, little sensitivity to temperature and relative humidity conditions, high oxidizing power, and the lack of formation of secondary undesired products \cite{a_adnew_gas-phase_2016, liu_photochemical_2011}.

An alternative to \ch{H2O2} as the primary oxidant for gas-phase AOPs is \ch{O3}. This has the advantage of directly oxidizing \ch{NO} while indirectly oxidizing \ch{SO2} when exposed to UV light. Some studies that dealt with industrial pollution removal in gas-phase AOPs with \ch{O3} include the work of Montecchio \textit{et al.} \cite{montecchio_development_2018, montecchio_development_2018} for volatile organic compounds (VOCs) and the work of the research group of M. Johnson \cite{johnson_gas-phase_2014,a_adnew_gas-phase_2016, meusinger_treatment_2017} for reduced sulfur compounds. However, to the best of the authors' knowledge, there are no studies that investigate the potential for radical oxidation of \ch{SO2} with \ch{O3} via UV light in the context of multi-pollutant removal solutions for cooperative denitration and desulfurization. Hence, in the present work numerical simulations of \ch{SO2} abatement by \ch{O3} and UV light in an idealized chemical reactor will be performed. In the following sections, details of the modeling approach will be provided along with the results obtained and a discussion of the conclusions that can be drawn from them.

\subsection*{2. Methods for simulation and analysis}
\addcontentsline{toc}{subsection}{2. Methods for simulation and analysis}

As described by Ducoste and Alpert \cite{ducoste_computational_2015}, simulating a UV-driven AOP system requires modeling of three different physical aspects that are interconnected. These are the modeling of fluid dynamics, chemical kinetics, and UV light distribution. In the following, the implementation of the modeling of each element will be described individually after presenting the reactor configuration that will be modeled.

\subsubsection*{2.1. Reactor configuration and conditions}
\addcontentsline{toc}{subsubsection}{2.1. Reactor configuration and conditions}

\begin{figure}[h!]
\centering
        \includegraphics[height=2in]{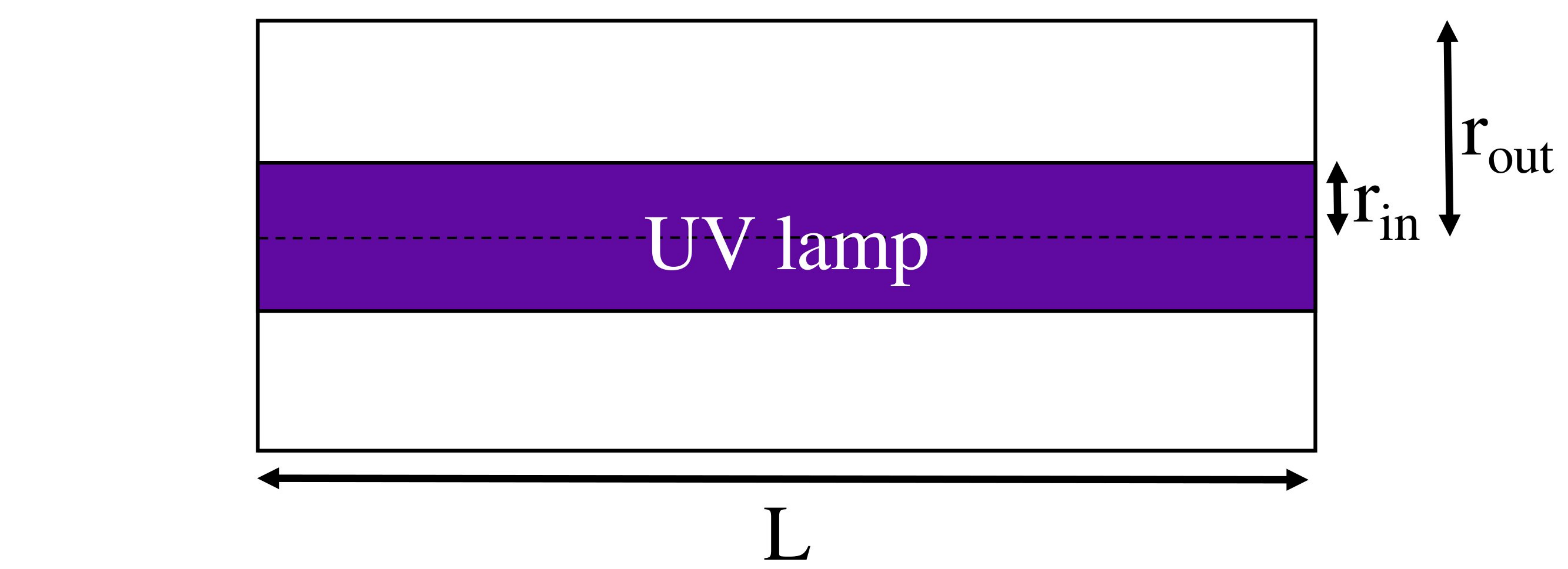}
        \caption{Sketch of the two-dimensional cross-section of the annular UV reactor employed. Not to scale.}
        \label{fig:2d}
\end{figure}

The geometry of the modeled reactor is an annular cylinder where the reactor and the lamp have the same length. Dimensions for it have been taken from Montecchio \textit{et al.} \cite{montecchio_development_2018}. Fig \ref{fig:2d} shows an illustration of the cross-section of the reactor. In the present study, the effect of reactor geometry has been studied. To do so, as in the work by Montecchio \textit{et al.} \cite{montecchio_development_2018} three different reactors have been analyzed. All three reactors have the same conditions except their outer radius varies. The relevant conditions for the three reactors studied, including the geometry and thermodynamic parameters, are presented in Table \ref{tab:params}.

\begin{table}[!htb]
    \caption{Parameters and conditions for the simulated CSTR reactor. All conditions are adapted from Montecchio \textit{et al.} \cite{montecchio_development_2018}. }
    \centering
    \begin{tabular}{p{6.5 cm} c c c}
        \hline \hline
Parameter                              & Symbol                 & Value  & Units \\ \hline
Lamp and reactor length                & $L$                    & 0.28   & m     \\
Lamp sleeve (inner radius)             & $r_{in}$               & 0.0115 & m     \\
Small reactor radius                   & $r_{A,out}$            & 0.0425 & m     \\
Medium reactor radius                  & $r_{B,out}$            & 0.0675 & m     \\
Big reactor radius                     & $r_{C,out}$            & 0.0820 & m     \\
Low-pressure UV lamp power             & $P$                    & 17     & W     \\
UV lamp efficiency at $\lambda=254$ nm & $\eta_{\lambda=254}$   & 0.33   & -     \\
Operating temperature                  & $T$                    & 298    & K     \\
Operating pressure                     & $p$                    & 1      & atm   \\ \hline \hline
    \end{tabular}
    \label{tab:params}
\end{table}

\subsubsection*{2.2. Reactor modeling}
\addcontentsline{toc}{subsubsection}{2.2. Reactor modeling}

In the present modeling approach, the fluid dynamics of the reactor were simplified to consider ideal mixing. By doing so, the concentrations of all species are assumed to be homogeneous throughout the reactor, often referred to as a continuously stirred tank reactor (CSTR). This is done to reduce the overall complexity of the implementation, which in turn reduces both the number of unknown variables in the model and the time required to produce results. This is advantageous in order to evaluate the sensitivity of the model to different initial conditions, geometries, chemical kinetics, and UV models. Furthermore, although the mixing is modeled as perfect, this approach provides an upper bound to what can be expected in more complex simulations and experimental studies.

In this work, the CSTR model as implemented in Cantera was employed \cite{goodwin_cantera_2016}. Cantera is an open-source package that aids in solving numerical chemistry problems and can be executed in Python 3 \cite{van_rossum_python_2009}. This software was used to solve the time-dependent equations that govern the evolution of the chemical species and the thermodynamic state of the reactor. 

\begin{figure}[h!]
\centering
        \includegraphics[height=2.5in]{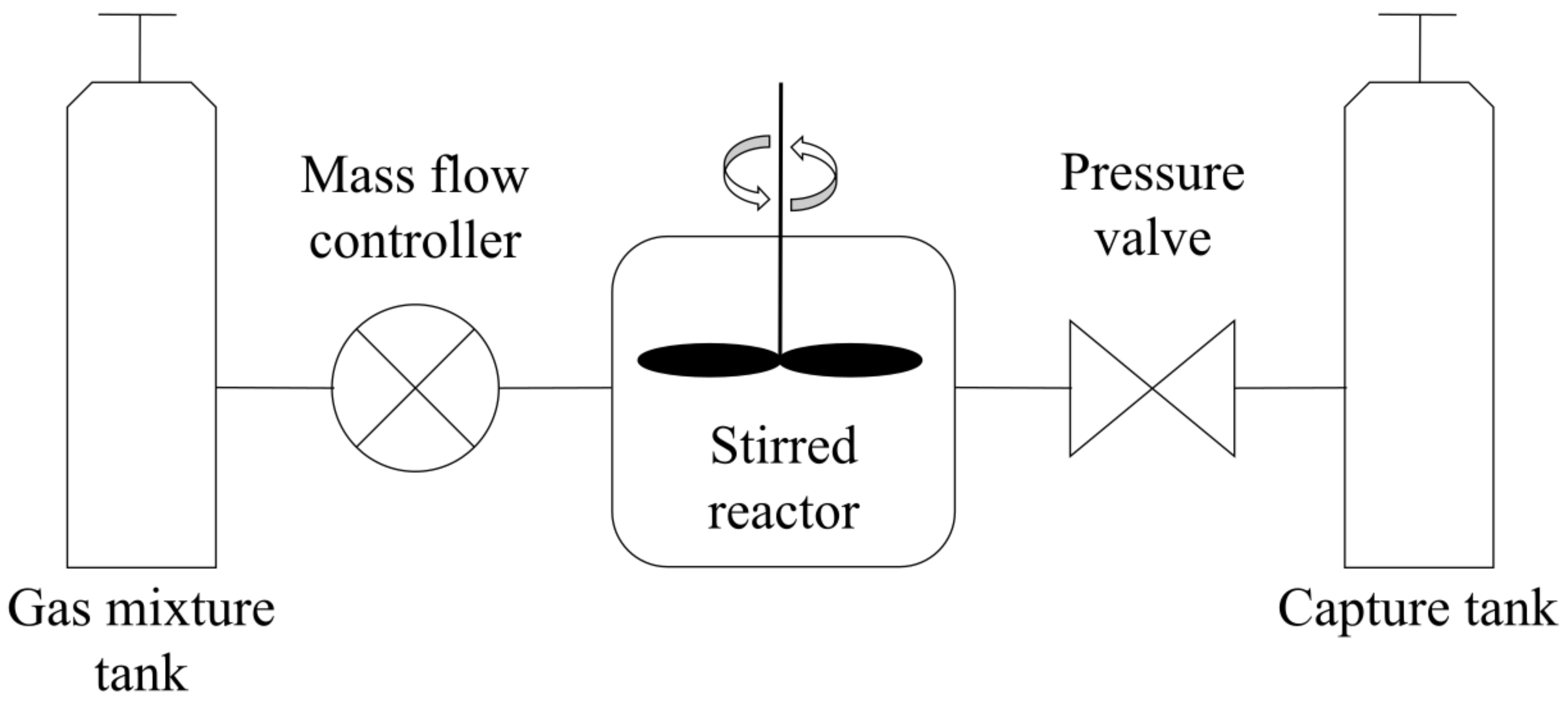}
        \caption{Illustration of the arrangement simulated for the CSTR reactor in Cantera \cite{goodwin_cantera_2016}.}
        \label{fig:cstr}
\end{figure}

A sketch of the reactor network simulated can be seen in Fig. \ref{fig:cstr}. Firstly, an infinite reservoir of the gas mixture is defined. The composition of the mixture tank is defined by the user by setting the initial conditions of the molar fraction of each of the species that are desired to react together. This mixture is considered perfectly mixed and is simulated overall as a perfect gas. In the present case, the simplest simulation included only air (as \ch{N2} and \ch{O2}), \ch{SO2}, \ch{O3} and water vapour (\ch{H2O}). Initial temperature and pressure conditions also have to be defined at this stage and were set to 295 K and 1 atm. Upstream of the tank a mass flow controller object is defined, which is fully described by the user-defined residence time once the volume of the reactor tank is fixed. It is then in the reactor object where the time loop is simulated until the user-defined maximum simulation time. Only here are reactions allowed to be carried out and hence composition changes with time. Downstream of the reactor, a pressure valve is simulated by setting a pressure valve coefficient which defines the efficiency to maintain the defined reactor pressure. Finally, a capture tank object is defined at the end of the network to store the resulting species from the reactions occurring in the stirred reactor.

\subsubsection*{2.3. Chemical kinetic modeling}
\addcontentsline{toc}{subsubsection}{2.3. Chemical kinetic modeling}

The information regarding the reactions that occur in the gas mixture in Cantera is stored in a kinetic mechanism. These files are the main chemical input to the model and have to be in the pre-defined Cantera format (CTI). The mechanism file stores elements and species of the gas as well as the thermodynamic properties of each specie. Following this, the chemical equations for the reactions between all defined species are provided. Each reaction is then modeled with a modified Arrhenius equation and thus the values for the pre-exponential factor $A$, the exponential factor $n$, and activation energy for the reaction $E_a$ must be provided. This allows the Cantera to numerically solve the system of ordinary differential equations (ODEs) describing the reaction mechanism in order to determine reaction rate coefficients. 

\begin{table}[!htb]
    \caption{Reduced kinetic mechanism based on the mechanism proposed by Meusinger \textit{et al.} \cite{meusinger_treatment_2017}. Here the units of the pre-exponential factor $A$ depend on the order the reaction. It is [$\mathrm{s}^{-1}$] for first order reactions, [$\mathrm{cm}^3 \; \mathrm{molecule}^{-1} \; \mathrm{s}^{-1}$] for second order reactions and  [$\mathrm{cm}^6 \; \mathrm{molecule}^{-2} \; \mathrm{s}^{-1}$] for third order reaction. R13 represents the photolysis of ozone by UV light. }
    \centering
    \begin{tabular}{c p{7cm} c c c}
        \hline \hline
ID & Reaction                      & $A$      & $n$  & $E_a/\mathrm{R}$ [K] \\ \hline
R1 & \ch{O + O3 -> 2 O2}            & 8.00E-12 & 0    & 2060             \\                      
R2 & \ch{O(^1D) + H2O -> 2 ^.OH}     & 1.63E-10 & 0    & -65              \\              
R3 & \ch{O(^1D) + N2 -> O + N2}    & 1.79E-11 & 0    & -110             \\                      
R4 & \ch{O + ^.OH -> O2 + H}         & 2.40E-11 & 0    & -110             \\                      
R5 & \ch{O + HO2 -> ^.OH + O2}       & 2.70E-11 & 0    & -224             \\                      
R6 & \ch{H + O2 + M -> HO2 + M}    & 2.17E-29 & -1.1 & 0                \\              
R7 & \ch{^.OH + O3 -> HO2 + O2}      & 1.70E-12 & 0    & 940              \\              
R8 & \ch{^.OH + HO2 -> H2O + O2}     & 4.80E-11 & 0    & -250             \\                      
R9 & \ch{^.OH + SO2 -> HSO3}         & 7.07E-11 & -0.7 & 0                \\              
R10 & \ch{O + SO2 + M -> SO3 + M}  & 4.00E-32 & 0    & 1000             \\                      
R11 & \ch{SO3 + H2O -> H2SO4}      & 1.20E-15 & 0    & 0                \\              
R12 & \ch{HSO3 + O2 -> HO2 + SO3}  & 1.30E-12 & 0    & 330              \\ 
R13 & \ch{O3 ->[h$\nu$] O(^1D) + O2 } & $k_{\mathrm{UV}}$   & 0    & 0              \\ \hline \hline 
    \end{tabular}
    \label{tab:mech}
\end{table}

The kinetic mechanisms employed in the present work was originally developed by Meusinger \textit{et al.} \cite{meusinger_treatment_2017}. This mechanism employs only the most relevant reactions among the 150 reactions provided, as predicted by their model. This reduced mechanism has a total of 13 reactions. The reactions and Arrhenius constants for the reduced mechanism are presented in Table \ref{tab:mech}.

Reactions R1-R12 presented in Table \ref{tab:mech} are thermal reactions while R13 is a photolytic reaction. Although the reaction rate of thermal reactions can be modeled by a modified Arrhenius equation, photolysis reactions cannot be generally modeled in the same way. If we consider a general photochemical reaction (\ch{A ->[h$\nu$] B}), this can decomposed into three steps \cite{aillet_photochemical_2013}. Here, $h$ is the Planck constant ($6.626 \times 10^{-34}\;\mathrm{J\:s\:photon^{-1}}$) and $\nu$ is the frequency of the radiated UV light. Hence, $Q_{\lambda}=h\nu$ is the energy of one photon at a given wavelength. Firstly an activation step (\ch{A ->[h$\nu$] A^*}) occurs when the initial ground state species absorbs a photon and produces an electronically excited version of the original specie. Once this activate species is created it can take two different paths. A deactivation step (\ch{A^* -> A}) occurs when the additional energy from the photon absorption is lost (by either a radiative mechanism such as fluorescence or by a non-radiative mechanism such as heat loss) and the activate specie returns to its original ground state. In the context of photo-activated reactions, this is often a negative outcome as it reduces the efficiency of the overall photochemical reaction. Finally, a reaction step (\ch{A^* -> B}) takes place when the excited species is converted to the desired photochemical product. 

The reaction rate for the photochemical activation step $r^{\lambda}_{(\ch{A -> A^*})}$ is directly proportional to the photon energy absorbed by the specie \cite{aillet_photochemical_2013}. Here the subscript described which reaction the reaction rate is referring to and the superscript $\lambda$ indicates the wavelength for which the photochemical reaction takes place (usually given in $\mathrm{nm}$).  The expression for the reaction rate of the photochemical activation step for species \ch{A} is at a certain wavelength $\lambda$ is:

\begin{equation}
    r^{\lambda}_{(\ch{A -> A^*})} = e_{\lambda,A}^{\prime} = \frac{\lambda}{N_A h c} e_{\lambda,A} \label{eq:1}
\end{equation}

\noindent{where $c$ is the speed of light ($2.998 \times 10^{8}\;\mathrm{m\:s^{-1}}$), $N_A$ is the Avogadro number ($6.023 \times 10^{23}\;\mathrm{mol^{-1}}$). The photon-based effective energy absorbed by \ch{A} is $e_{\lambda,A}^{\prime}$ and is often referred to as the local volumetric rate of photon absorption or LVRPA in $\mathrm{{einstein}\:s^{-1}\:m^{-3}}$ which is equivalent to $\mathrm{mol}$-$\mathrm{photon} \mathrm{\:s^{-1} \:m^{-3}}$ \cite{cassano_photoreactor_1995}. This can be converted to the energy-based effective energy absorbed by \ch{A}, $e_{\lambda,A}$, also known as the local volumetric rate of energy absorption or LVREA (in $\mathrm{{W}\:m^{-3}}$) by multiplying $e_{\lambda,A}^{\prime}$ by the energy contained in one mole of photons at a given wavelength (i.e. $Q_{\lambda,\mathrm{mol}}=N_A h \frac{c}{\lambda}$)}.

For the deactivation and reaction steps, their respective reaction rates can be formulated by considering both as first order thermal reactions:

\begin{equation}
    r_{(\ch{A^* -> A})} = k_{(\ch{A^* -> A})} [\ch{A^*}] \label{eq:2}
\end{equation}

\begin{equation}
    r_{(\ch{A^* -> B})} = k_{(\ch{A^* -> B})} [\ch{A^*}] \label{eq:3}
\end{equation}\label{}

\noindent{where $k_{(\ch{A^* -> A})}$ and $k_{(\ch{A^* -> B})}$ are the reaction rate constants for the deactivation and reaction step, respectively and $[\ch{A^*}]$ is the concentration of species \ch{A^*} in $\mathrm{{mol}\:m^{-3}}$. Putting Eqs \ref{eq:1} - \ref{eq:3} together the time evolution of the concentration of species \ch{A^*} can be written as:}

\begin{equation}
    \dfrac{d [\ch{A^*}]}{dt} = r^{\lambda}_{(\ch{A -> A^*})} - r_{(\ch{A^* -> A})} - r_{(\ch{A^* -> B})}
\end{equation}

\noindent{It can be assumed that the concentration of the photo-activated intermediate species \ch{A^*} reaches a quasi-steady state and so $\dfrac{d [\ch{A^*}]}{dt} \approx 0$ \cite{aillet_photochemical_2013}. Therefore, the quasi-steady state concentration of species \ch{A^*} can be expressed as:}

\begin{equation}
    [\ch{A^*}] = \dfrac{e_{\lambda,A}^{\prime}}{k_{(\ch{A^* -> A})} + k_{(\ch{A^* -> B})}} 
\end{equation}

\noindent{Therefore, the rate of consumption of \ch{A}, $r_A$, which equals the rate of formation of \ch{B}, $r_B$, can be written for the overall photochemical reaction as:}

\begin{equation}
     - r_A = r_B = k_{(\ch{A^* -> B})} [\ch{A^*}] = \dfrac{k_{(\ch{A^* -> B})}}{k_{(\ch{A^* -> A})} + k_{(\ch{A^* -> B})}} e_{\lambda,A}^{\prime} = \phi_{\lambda}  e_{\lambda,A}^{\prime} \label{eq:6}
\end{equation}

\noindent{where $\phi_{\lambda}$ is the quantum yield for the overall reaction at a given wavelength.}

Another description for the quantum yield can be obtained by defining the overall photochemical reaction rate in such a way that it satisfies the first and second laws of photochemistry \cite{bolton_rethinking_2015}. The first law states that it is only the light absorbed by a species which can effectively produce a photochemical change. This law is also known as the Grotthus-Draper law. The second law, known as the Stark-Einstein law, says that each photon can at most cause photochemical change of one single molecule which has absorbed it. Therefore, it defines that the primary quantum yields of a molecule must sum one. Hence, the quantum yield itself is defined as the number of molecules of the reactant species \ch{A} consumed per unit time, $n(A)$,  over the total number of photons absorbed $\Phi_p^{abs}$ \cite{t_oppenlander_photochemical_2003}:

\begin{equation}
    \phi_{\lambda} = \dfrac{dn(\ch{A})/dt}{\Phi_p^{abs}}
\end{equation}

\noindent{Converting the number of molecules of \ch{A}, $n(\ch{A})$, to the concentration [\ch{A}] by \cite{t_oppenlander_photochemical_2003}:}

\begin{equation}
    dn(\ch{A}) = [\ch{A}] N_A dV
\end{equation}

\noindent{where $dV$ is an infinitesimally small volume element that contain $dn$ molecules of \ch{A}. Rearranging, one can write an expression for the consumption of \ch{A} \cite{t_oppenlander_photochemical_2003}:}

\begin{equation}
    -r_A = -\dfrac{d [\ch{A}]}{dt} = \phi_{\lambda} \dfrac{\Phi_p^{abs}}{N_A V} = \phi_{\lambda} e_{\lambda,A}^{\prime} = \phi_{\lambda} \frac{\lambda}{N_A h c} e_{\lambda,A} \label{eq:9}
\end{equation}

\noindent{for which we obtain the same expression as Eq. \ref{eq:6}. This is the equation that will allow expressing the photochemical reaction kinetics. However, first, the local volumetric rate of energy absorption $e_{\lambda,A}$ must be obtained. This will be discussed in Section 2.3. Note that as the reactor is assumed to be perfectly mixed, the reaction rate will only depend on time assuming fixed geometry and constant lamp power output. Hence, despite Eq. \ref{eq:9} being also applicable to any individual point in the reactor, for the present work the reaction rates will be defined for the whole volume. For this to be achieved, the LVRPA or LVREA will be required to be volume-averaged quantities as the remainder will no have a spatial dependency. Defining this explicitly, the rate of consumption of \ch{A} for the present CSTR reactor is defined as:}

\begin{equation}
    -\dfrac{d [\ch{A}]}{dt} = \phi_{\lambda} \langle e_{\lambda,A}^{\prime} \rangle_V = \phi_{\lambda} \frac{\lambda}{N_A h c} \langle e_{\lambda,A} \rangle_V \label{eq:10}
\end{equation}

\subsubsection*{2.4. UV distribution modeling}
\addcontentsline{toc}{subsubsection}{2.4. UV distribution modeling}

In a general annular photoreactor where we can assume that the characteristic length is much larger than the UV radiation wavelength, the general transport equation for single-wavelength radiations, also known as the radiation transport equation (RTE) reads as follows \cite{modest_radiative_2013, huang_evaluation_2011}:

\begin{equation}
    \underbrace{\dfrac{1}{c}\dfrac{\partial I_{\lambda}}{\partial t}\vphantom{\int_{4 \pi}}}_{\text{Transient}} + \underbrace{\dfrac{\partial I_{\lambda}}{\partial s}\vphantom{\int_{4 \pi}}}_{\text{Along $s$}} = \underbrace{j_{\lambda}\vphantom{\int_{4 \pi}}}_{\text{Emission}} - \underbrace{\alpha_{\lambda} I_{\lambda} \vphantom{\int_{4 \pi}}}_{\text{Absorption}} - \underbrace{\sigma_{\lambda,s} I_{\lambda} \vphantom{\int_{4 \pi}}}_{\text{Out-scattering}} + \underbrace{\dfrac{\sigma_{\lambda,s}}{4 \pi} \int_{4 \pi} I_{\lambda}(\textbf{{\^{s}}}_i) \varphi_{\lambda} (\textbf{{\^{s}}}_i, \textbf{{\^{s}}}) d\Omega_i}_{\text{In-scattering}}
\end{equation}

\noindent{where $I_{\lambda}$ is the radiation intensity in $\mathrm{{einstein}\:m^{-2}\:s^{-1}}$ at a given radiation wavelength $\lambda$, $s$ is the radiation ray path and $\alpha_{\lambda}$ is the (Naperian) absorption coefficient of the medium in $\mathrm{m^{-1}}$. A thorough description of the remainder terms can be found in Modest \textit{et al.} \cite{modest_radiative_2013}. Given that the boundary conditions are steady, the ration field will instantaneously reach a steady-state and the transient term can be neglected \cite{huang_evaluation_2011}. This is valid only when the speed of light is much larger than the ratio between the characteristic length and the characteristic time. Nevertheless, this is the case in most engineering applications \cite{modest_radiative_2013}. In UV reactors, temperatures are generally sufficiently low to assume that the emission from black-body radiation is negligible \cite{pareek_light_2008}. Hence, the emission term can be neglected. Finally, as the considered medium (i.e. air or water) does not have a large concentration of solid particles or scattering can be generally assumed to be non-elastic, the in- and out-scattering terms can be neglected \cite{elyasi_general_2010}. This simplification is also known as the purely absorbing medium \cite{huang_evaluation_2011}. Therefore, the final expression for the RTE when all relevant assumptions are made reads:}

\begin{equation}
    \dfrac{\partial I_{\lambda}}{\partial s} + \alpha_{\lambda} I_{\lambda} = 0
\end{equation}

\noindent{which, when integrated is:}

\begin{equation}
    I_{\lambda} = I_{\lambda,0} \: \exp \left({-\int_{0}^{l} \alpha_{\lambda} ds}\right) \label{eq:13}
\end{equation}

\noindent{where $I_{\lambda,0}$ is the radiation intensity without absorption and $l$ is the distance from the light radiation source to the point at which the radiation intensity  $I_{\lambda}$ is going to be computed. Applying the definition of the absorption coefficient for the medium (i.e. the mixture of species) and the additive property Eq. \ref{eq:13} can be expressed as \cite{t_oppenlander_photochemical_2003}:}

\begin{equation}
    I_{\lambda} = I_{\lambda,0} \: \exp \left({-\int_{0}^{l} \sum_{i=1}^{N} \varepsilon_{\lambda,i} [\:i\:] \ln 10 \: ds}\right)
\end{equation}

\noindent{where $i$ is each individual species in the medium, $N$ is the total number of species contained in the medium, $[\:i\:]$ is the concentration of the i-\textit{th} specie and $\varepsilon_{\lambda,i}$ is the wavelength dependent molar decadic absorption coefficient in $\mathrm{m^2\:{mol}^{-1}}$ of the i-\textit{th} specie. The $\ln 10$ term is required to convert from the Naperian absorption coefficient $\alpha_{\lambda}$ to the molar decadic absorption coefficient $\varepsilon_{\lambda,i}$ \cite{t_oppenlander_photochemical_2003}. Assuming that the the molar absorption coefficient is path-independent and the concentration of each species is homogeneous because the medium perfectly mixed:}

\begin{equation}
    I_{\lambda} = I_{\lambda,0} \: \exp \left(\ln 10 \: l \sum_{i=1}^{N} \varepsilon_{i,\lambda} [\:i\:] \right) \label{eq:15}
\end{equation}

In this form the simplified RTE in Eq. \ref{eq:15} is often referred to as the Beer-Lambert Law. This simple equation describes how the different species present in the medium through which a radiation ray is traveling partially absorb its energy to reduce the incoming intensity. Eq. 14 is suited for the description of the radiation intensity at a single point in the annular reactor domain.

The LVRPA for species \ch{A} which undergoes photochemical change can be expressed as \cite{coenen_modeling_2013,cassano_design_2005,cassano_photoreactor_1995}:

\begin{equation}
    e_{\lambda,A}^{\prime} = \alpha_{\lambda,A} I_{\lambda}
\end{equation}

\noindent{substituting in Eq. \ref{eq:15} and expressing the Naperian absorption coefficient of species \ch{A} in terms of the molar decadic absorption coefficient and the concentration of species \ch{A} the LVRPA equation reads:}

\begin{equation}
    e_{\lambda,A}^{\prime}(\mathbf{x},t) = \ln 10 \: \varepsilon_{A,\lambda} [\ch{A}(t)] I_{\lambda,0}(\mathbf{x}) \: \exp \left(\ln 10 \: l(\mathbf{x}) \sum_{i=1}^{N} \varepsilon_{i,\lambda} [i(t)] \right) \label{eq:17}
\end{equation}

\noindent{where the dependencies of each term with respect to space (described as a spatial vector $\mathbf{x}$) and time $t$) have been made explicit. As was the case for Eq. \ref{eq:15}, Eq. \ref{eq:17} describes the LVRPA for a single point in the reactor. To calculate the LVRPA for the whole reactor, the hollow cylinder volume was discretized. This will be discussed in detail in Section 3. For each mesh element, the radiation intensity $I_{\lambda}$ was computed taking the value calculated at the cell center. Then the volume-averaged radiation intensity is computed as:}

\begin{equation}
    \langle I_{\lambda} \rangle_V = \dfrac{\sum\limits_{j=1}^{M} V_j I_{j,\lambda}}{\sum\limits_{j=1}^{M} V_j} \label{eq:18}
\end{equation}

\noindent{where $M$ is the total number of cell elements, $V_j$ is the volume of the $j$-th element and $I_{j,\lambda}$ is the radiation intensity of the $j$-th element. Therefore, the equation that described the volume-averaged LVRPA reads:}

\begin{equation}
    \langle e_{\lambda,A}^{\prime} \rangle_V = \ln 10 \: \varepsilon_{A,\lambda} [\ch{A}] \langle I_{\lambda} \rangle_V \label{eq:19}
\end{equation}

\noindent{If Eq. \ref{eq:19} is substituted into Eq. \ref{eq:10} we obtain a pseudo-first order reaction rate equation for the photochemical reaction \ch{A ->[h$\nu$] B}:}

\begin{equation}
    -\dfrac{d [\ch{A}]}{dt} = \phi_{\lambda} \ln 10 \: \varepsilon_{A,\lambda} [\ch{A}] \langle I_{\lambda} \rangle_V = k_{\mathrm{UV}} [\ch{A}]
\end{equation}

\noindent{where $k_{\mathrm{UV}}$ is a function of the geometry of the reactor and the discretization of it, the UV lamp characteristics and the radiation model employed and the concentrations of the different radiation-absorbing species (including $\ch{A}$) which are themselves a function of time.}

The only remaining unknown is $I_{\lambda,0}$. This will be calculated using several possible radiation models. Generally, these models can be divided into incidence and emission models. The former assumes a radiation distribution inside the reactor while the latter models the type of radiation emitted by the source. Usually emission models are preferred as incidence models require experimental parameters \cite{pareek_light_2008}. Comprehensive reviews of these models can be found elsewhere in the literature \cite{pareek_light_2008, liu_evaluation_2004,alfano_radiation_1986} so only the models employed in the present work will be described briefly. All models presented in the following are known as line source models. Here, the UV lamp emission is modeled as coming from a one-dimensional line. Other models that simulate the UV lamp as a two-dimensional surface or a three-dimensional volume also exist but the line source models are simpler and offer reasonable accuracy \cite{pareek_light_2008}. The nomenclature for these models is not consistent throughout the literature, however, in the present work the terminology employed by Liu \textit{et al.} \cite{liu_evaluation_2004} will be used as their application (i.e. AOPs for water treatment) is closer to the intended application of this study. Note that the following models are usually presented in the literature including absorption (i.e. $I_{\lambda}$ instead of $I_{\lambda,0}$) but will be described in their simpler form as absorption will be treated separately. For all cases, the origin of the coordinate system will be placed at the center of the UV lamp. These models and their equations are:

\begin{itemize}
    \item Radial model (RAD): the radial model is the most simple radiation model and was first introduced by Harris and Dranoff \cite{harris_study_1965}. This model assumes that the light source is only emitting UV radiation in the radial direction and hence can be considered a one-dimensional model. Mathematically, for any point in the domain, this reads: 
    
    \begin{equation}
        I_{\lambda,0}(r) = \frac{P_{\lambda}^{\prime}}{2 \pi r L}
    \end{equation}
    
    \noindent{where $r$ is the radial coordinate (i.e. the radial distance from the lamp center to a specific point inside the reactor) $L$ is the length of the lamp and $P_{\lambda}^{\prime}$ is the useful UV lamp power emitted a specific wavelength $\lambda$ in $\mathrm{einstein \: s^{-1}}$. Note that $P_{\lambda}^{\prime}$ can be obtained from $P_{\lambda}$, the useful UV lamp power emitted a specific wavelength $\lambda$ in $\mathrm{W}$, as:}
    
    \begin{equation}
        P_{\lambda}^{\prime} = Q_{\lambda,\mathrm{mol}} P_{\lambda} = Q_{\lambda,\mathrm{mol}} P \eta_{\lambda} 
    \end{equation}
    
    where $P$ is the UV lamp power in $\mathrm{W}$ and $\eta_{\lambda}$ is the UV lamp efficiency at a specific wavelength $\lambda$.
    
    \item Multiple points source summation (MPSS): originally introduced by Jacob and Dranoff \cite{jacobm_light_1970}, the MPSS model divides the lamp into several equispaced point light sources along the lamp centerline. These light sources, unlike in the RAD model, are not limited to emitting light in a single direction, but do so in all available directions. This is often referred to as spherical emission. As in an annular photoreactor, there is azimuthal symmetry, this model is two-dimensional. The radiation intensity without absorption from one single point light source $k$ to a single point in the domain is:
    
    \begin{equation}
        l_{k,\lambda,0}(r,z) = \frac{P_{\lambda}^{\prime}/N_{ls}}{4 \pi \rho^2}
    \end{equation}
    
    \noindent{where $z$ is the axial coordinate (i.e. the axial distance from the lamp center to a specific point inside the reactor), $N_{ls}$ is the total number of light sources and $\rho$ can be described as:}
    
    \begin{equation}
        \rho = \sqrt{r^2+(z-h)^2}
    \end{equation}
    
    \noindent{where $h$ is the axial distance from the lamp center to the point light source. Then the total radiation intensity (without considering absorption) for any point in the domain is:}
    
    \begin{equation}
        I_{\lambda,0}(r,z) = \sum\limits_{k=1}^{N_{ls}} l_{k,0,\lambda}(r,z) \label{eq:25}
    \end{equation}
    
    The MPSS model introduces the unknown parameter of the total number of light sources with which to discretize the UV lamp. This parameter will be addressed in Section 3.
    
    \item Multiple segment source summation (MSSS): the MSSS model was an extension of the MPSS model introduced by Bolton (as cited by Liu \textit{et al.} \cite{liu_evaluation_2004}). Here, the apparent overprediction of the MPSS model was tackled by assuming that the light emission from the point light sources was diffuse rather than spherical. This amounts to considering the point light sources as finite cylindrical segments. This reads:
    
    \begin{equation}
        l_{k,\lambda,0}(r,z) = \frac{P_{\lambda}^{\prime}/N_{ls}}{4 \pi \rho^2} \cos \left( \theta \right)
    \end{equation}
    
    where $\theta$ is defined as the angle formed between the radial direction and the direction that connects the point in the domain at which the radiation intensity is being calculated with the $k$-th point light source:
    
    \begin{equation}
        \theta = \arctan{\dfrac{|z-h|}{r}}
    \end{equation}
    
    Then, the total radiation intensity is computed following Eq. \ref{eq:25}.
    
    \item Line source integration (LSI): the LSI model was first introduced by Jacob and Dranoff \cite{jacobm_light_1970} and later assessed by Blatchley \cite{blatchley_numerical_1997} for collimated-beam AOP reactors. Formally, this method is equivalent to the MPSS with an infinite number of discrete point light sources. Therefore, it is the integral version of the MPSS. Without considering the absorption of the medium, the LSI model has an analytical solution:
    
    \begin{equation}
        I_{\lambda,0}(r,z) = \frac{P_{\lambda}^{\prime}}{4 \pi r L} \left( \arctan{\frac{L/2+z}{r}} + \arctan{\frac{L/2-z}{r}} \right)
    \end{equation}
    
    It should be noted that when this model is employed all absorption is neglected. This includes the absorption of the photo-activated species which is therefore not physical. Although some efforts have been made to include absorption terms \textit{a posteriori} (i.e. after integration) \cite{montecchio_development_2018} experimental fitting is required. This defeats the purpose of \textit{a priori} modeling and hence the LSI model will only be employed for evaluating the value of the MPSS model with an infinite number of point light sources. 
    
    \item Modified LSI (RAD-LSI): the RAD-LSI model was proposed by Liu \textit{et al.} and incorporates both the RAD and LSI models together. This is done because the LSI model is able to correctly predict far-field radiation intensities but not the near-field distribution. This happens because, when approaching the UV lamp, most radiation incident on a point will come from the light source immediately close to it. Hence, close to the lamp the radiation can be through as being is mostly radial and so the RAD model will better predict the near field. Formally this is described as:
    
    \begin{dmath}
            I_{\lambda,0}(r,z) = \mathrm{min}\left[\mathrm{RAD},\;\mathrm{LSI} \right] = \mathrm{min}\left[ \frac{P_{\lambda}^{\prime}}{2 \pi r L}, \; \frac{P_{\lambda}^{\prime}}{4 \pi r L} \left( \arctan{\frac{L/2+z}{r}} + \arctan{\frac{L/2-z}{r}} \right) \right]
    \end{dmath}
    
    where $\mathrm{min}\left[\mathrm{RAD},\;\mathrm{LSI} \right]$ is minimum function between the RAD and LSI models.
\end{itemize}

Often when discussing the models presented above in the context of water treatment AOPs, great importance is given to evaluate the effect of additional terms that change the radiation intensity distribution \cite{bolton_calculation_2000, liu_evaluation_2004}. These terms are mainly absorption, reflection, and refraction. For air, Liu \textit{et al.} argue that the latter two terms can be omitted \cite{liu_evaluation_2004}. Nevertheless, for cases with low absorption neglecting reflection could yield lower performance. The effect of the absorption of the medium should be evaluated separately because it will depend on the species present and the wavelength at which the reactor will be operating. As presented in Eq. \ref{eq:15}, the Beer-Lambert law describes the reduction in radiation intensity by an exponential term with accounts for the distance traveled by a UV light ray $l$, the concentration of the $i$-th species and how much that species absorbs light. The latter parameter $\varepsilon_{\lambda,i}$ is often reported in the literature indirectly as the (Naperian) absorption cross-section coefficient $\sigma_{\lambda,i}$ in $\mathrm{cm^{2}}$. To obtain the former from the latter Eq. \ref{eq:30} can be used:

\begin{equation}
    \varepsilon_{\lambda,i} = \frac{\sigma_{\lambda,i} N_A}{10^4 \ln{10}} \label{eq:30}
\end{equation}

\noindent{Note that Eq. \ref{eq:30} yields $\varepsilon_{\lambda,i}$ in S.I. units (i.e. $\mathrm{m^2\:{mol}^{-1}}$) but is only valid for $\sigma_{\lambda,i}$ in $\mathrm{cm^{2}}$ as it is in these units how this value is given in the literature.}

A valuable resource for obtaining accurate and updated values for the absorption cross-section coefficient of different species is the MPI-Mainz UV/VIS Spectral Atlas \cite{keller-rudek_mpi-mainz_2013}. Table \ref{tab:cross} presents the values for the absorption cross-section coefficients for the species in the kinetic mechanism as presented in Table \ref{tab:cross}. Specific references are mentioned when more than one option was available. The references employed were selected based on recency and/or if the publication was a review paper. The temperature and wavelength at which these values are evaluated are as close as possible to the operating conditions of the reactor modeled (i.e. 300 K and 254 nm).

\begin{table}[!htb]
    \caption{Absorption cross-section coefficients for different species present in the present reactor simulations. All values were obtained from the MPI-Mainz UV/VIS Spectral Atlas \cite{keller-rudek_mpi-mainz_2013} for the species in gas phase at $\sim300$ K and for UV radiation at a radiation wavelength of 254 nm, except otherwise stated. }
    \centering
    \begin{tabular}{c c c}
        \hline \hline
Species    & Absorbtion cross section, $\mathrm{cm^{2}}$ & Reference \\ \hline
\ch{O3}    & 1.132935E-17                                & Hodges \textit{et al.} \cite{hodges_recommendation_2019} \\    
\ch{O2}    & 1.477865E-24                                & Bogumil \textit{et al.} \cite{bogumil_measurements_2003} \\    
\ch{N2}    & Not reported above 150 nm                   & Spectral Atlas \cite{keller-rudek_mpi-mainz_2013} \\
\ch{SO2}   & 1.551258E-19                                & Bogumil \textit{et al.} \cite{bogumil_measurements_2003} \\
\ch{H2O}   & Approaches 0                                & Ranjan \textit{et al.} \cite{ranjan_photochemistry_2020} \\
\ch{HO2}   & 2.63E-19                                    & Tyndall \textit{et al.} \cite{tyndall_atmospheric_2001} \\
\ch{SO3}   & 1.34E-20                                    & Burkholder \textit{et al.} \cite{burkholder_uv_1997} \\
\ch{H2SO4} & 7.19E-22 (at 248 nm)                        & Farahani \textit{et al.} \cite{farahani_simulated_2019} \\ \hline \hline 
    \end{tabular}
    \label{tab:cross}
\end{table}

To simplify the computations, the contribution of some species to medium absorption was neglected. Species with very low or negligible absorption cross-section coefficients like \ch{N2}, \ch{H2O}, \ch{O2} or \ch{H2SO4} were not considered. Furthermore, the concentration of others such as \ch{HO2} and \ch{SO3} will be too low to have a meaningful impact. Hence only \ch{O3} and \ch{SO2} are considered as UV radiation-absorbing species for a wavelength of 254 nm.

\subsection*{3. Results and discussion}
\addcontentsline{toc}{subsection}{3. Results and discussion}

\subsubsection*{3.1. Grid convergence}
\addcontentsline{toc}{subsubsection}{3.1. Grid convergence}

One of the earliest decisions that must be taken to simulate the UV CSTR is how the discretization of the control volume will be performed. As has been presented in Section 2.4, all the radiation models employed are either one-dimensional or two-dimensional. This defines, at most, only two unknowns which are the number of cell elements along the radial coordinate, $n_r$,  and along the axial coordinate, $n_z$.

To evaluate which value for the total number of cells is high enough to ensure good accuracy for a simple simulation was performed using the LSI model. The small reactor radius is employed and both coordinates are discretized with the same number of points. Fig. \ref{fig:error} shows the convergence of the value for the volume-averaged radiation intensity (VARI) with increasing number of cells. Both the absolute value of the VARI as well as the relative rate of change (i.e. the absolute value of the change between one simulation and the next) is presented. As can be seen, above 100 cells there is little change in the VARI. This corresponds to 10 cells in each direction. Nevertheless, to ensure a value below 0.1\% for the relative rate of change of the VARI a the number of cells chosen for the subsequent studies is 225, which corresponds to 15 cells in each direction. When a larger reactor (i.e. larger outer radius) is employed the number of cells will be increased to maintain the ratio between the hollow cylinder radius and the number of cells in the radial direction constant.

\begin{figure}[h!]
\centering
        \includegraphics[height=6.1cm]{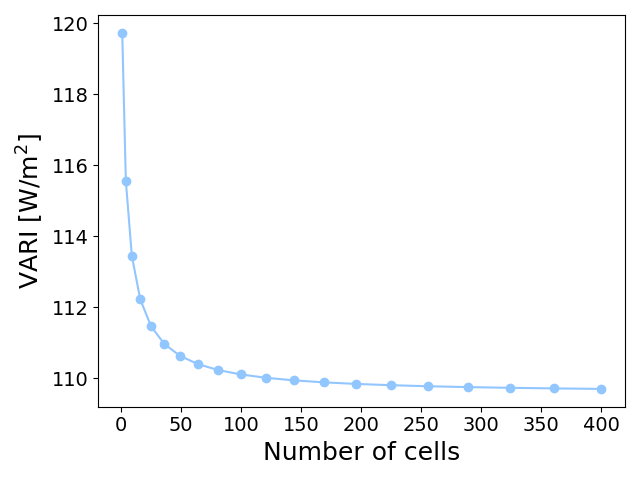}
        \includegraphics[height=6.1cm]{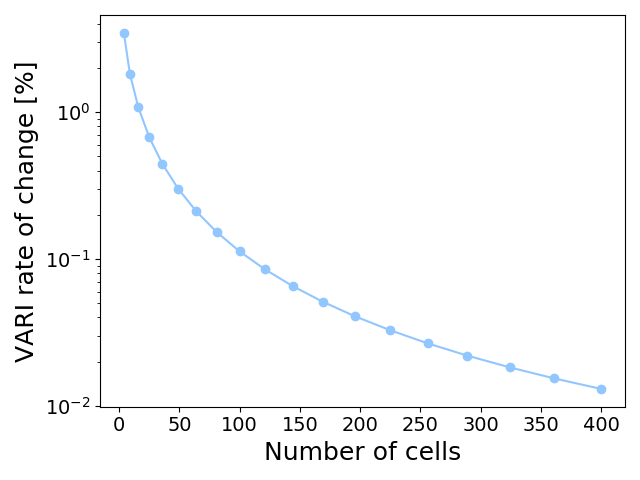}
        \caption{Evolution of the the volume averaged radiation intensity (VARI) for different total number of cells using the LSI method. \textit{Left}: total value of the VARI. \textit{Right}: relative rate of change of the VARI between successive simulations in \%.}
        \label{fig:mesh}
\end{figure}

\subsubsection*{3.2. Optimization of the number of point light sources}
\addcontentsline{toc}{subsubsection}{3.2. Optimization of the number of point light sources}

Although in the original work by Jacob and Dranoff \cite{jacobm_light_1970} less than $0.1\%$ error was achieved with only 10 light sources, a more recent work by Bolton \cite{bolton_calculation_2000} suggests that less than $1\%$ is usually achieved with close to 1000 discretization points. Current implementations of the MPSS model tend to use the recommendation of Bolton despite the large discrepancy between both values. Unfortunately, a large number of discrete point light sources leads to a considerable increase in the computation cost. Furthermore, if the radiation field is evaluated in each iteration of a time loop inside a reactor discretized with many cell elements even simple CSTR simulations can be costly.

To tackle this the defining property of the LSI model was employed. As the LSI model is the MPSS model without absorption in the limit when the number of discrete point light sources ($N_{ls}$) tends to infinity, a simple error equation can be written as:

\begin{equation}
    \epsilon = \frac{|I_{\lambda,0}^{\mathrm{MPSS}}-I_{\lambda,0}^{\mathrm{LSI}}|}{I_{\lambda,0}^{\mathrm{LSI}}}
\end{equation}

\begin{figure}[h!]
\centering
        \includegraphics[width=0.7\textwidth]{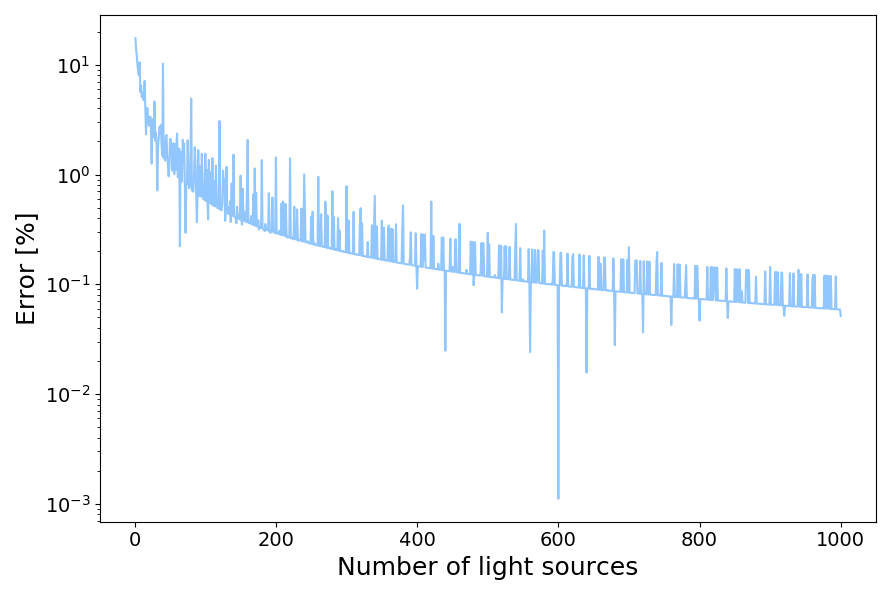}
        \caption{Evolution of the error for the MPSS model without considering the absorption of the medium (as compared to its integrated form, the LSI model) with the discrete number of light sources employed to model the UV lamp.}
        \label{fig:error}
\end{figure}

\noindent{where $\epsilon$ is the error incurred. Hence, for fixed conditions several values for $I_{\lambda,0}^{\mathrm{MPSS}}$ can be computed until the error is as small as a user-defined tolerance. Such computation has been implemented in the present work. The evolution of the error with the number of light sources can be seen in Fig. \ref{fig:error}. Here all the error incurred in for all integer values of $N_{ls}$ from 1 to 1000 is plotted. As can be observed, although a clear trendline is visible, there is considerable variability that appears unpredictable. Hence, although using a very large value for $N_{ls}\sim1000$ can yield low error for this specific configuration $\sim 0.1\%$, the same or even lower errors can be achieved by using much lower values. It can be seen that close to $N_{ls}=400$ two distinct values produce sharp valleys which yield substantially low errors. Therefore, finding the lowest possible value of $N_{ls}$ that produces the desired error is a worth-while time investment.}

Considering the large errors that are incurred in by some of the main assumptions in the current work (e.g. ideal mixing causing perfectly homogeneous concentrations) the target error for present computations was fixed to $1\%$. As a rule of thumb less than 50 points were required for all studied conditions in order to produce this error. In particular, for the 225 cell mesh, only 24 points were enough to obtain a 0.95\% error with respect to the same LSI simulation. The speedup achieved by this optimization was also tested. For all mesh sizes studied, the case with the computed optimum value of $N_{ls}$ was around 7 times faster than the case using 1000 points. This optimum value of $N_{ls}$ is also employed when using the MSSS method, however, the value is not computed using the MSSS method but rather by using the MPSS method and comparing it with the results from the LSI prediction.

\subsubsection*{3.3. Parametric study}
\addcontentsline{toc}{subsubsection}{3.3. Parametric study}

In the following subsection, the different parameters that make up the CSTR UV simulation as explained above will be evaluated. This will be done by selectively varying some of the different values and aiming to establish an optimum or ideal range within possible trade-offs. Some parameters that were found to not significantly affect the removal rate of \ch{SO2} were not shown here. These include the initial concentration of \ch{SO2}, the relative humidity (i.e. concentration of water vapor), and the gas temperature. All of these were tested within a reasonable rage starting from the baseline conditions of the simulation. The data shown for each simulation are obtained for the final simulation time when results are considered converged in time.  

\paragraph*{Radiation model.}

The five different radiation models presented in Section 2.4 have been compared under equal conditions. The reactor employed was the smallest of the three (see Table \ref{tab:params}) and was discretized with 225 mesh elements. For the MPSS and MSSS models, the previously obtained optimum value of $N_{ls}=24$ was employed. A total of 300 parts per million (ppm) of \ch{SO2} and 10,000 ppm of \ch{H2O} were employed. The concentration of \ch{O3} was varied to account for a range of molarity ratios (i.e. \ch{O3}/\ch{SO2}). A deliberately short residence time of 10 s was defined to better assess the differences between models.

\begin{figure}[h!]
\centering
        \includegraphics[height=6.1cm]{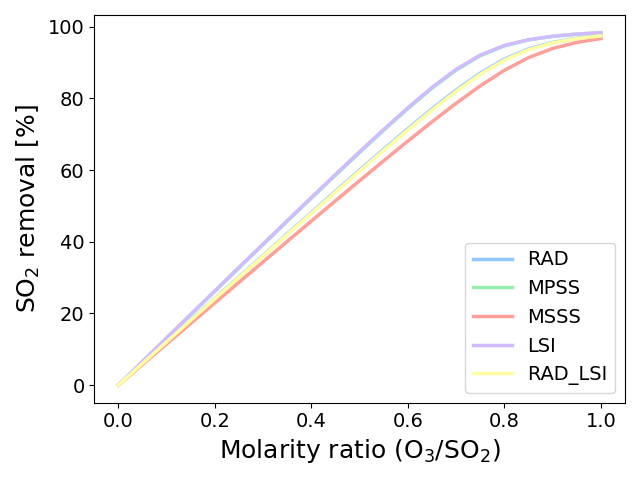}
        \includegraphics[height=6.1cm]{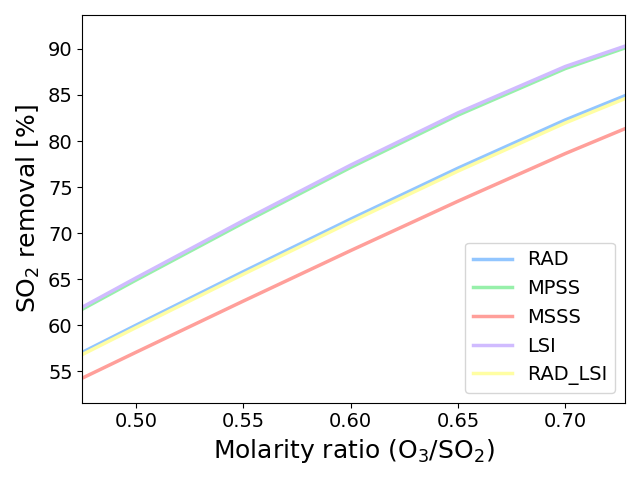}
        \caption{Evolution of the percentage removal of \ch{SO2} for different molarity ratios and different radiation models. \textit{Left}: Molarity ratio from 0 to 1. \textit{Right}: Zoomed view, molarity ratio from 0.5 to 0.7.}
        \label{fig:models}
\end{figure}

Fig. \ref{fig:models} shows the removal of \ch{SO2} by UV-\ch{O3} at different molarity ratios for the five models described. As expected the LSI model produces the highest \ch{SO2} removal as it is an idealized model without absorption. However, this absorption is low for the present case as only 300 ppm of \ch{SO2} and at most 300 ppm of \ch{O3} are actively absorbing UV radiation at $\lambda=254$ nm. Hence, the MPSS can be observed to perform similarly to the LSI under these conditions. This low light absorption for gas-phase UV-\ch{O3} systems was already reported by Montecchio \textit{et al.} \cite{montecchio_development_2018}. The RAD and RAD-LSI models also yield very similar results owing to the nature of the RAD-LSI formulation. However, the MSSS model provides a considerably lower prediction for \ch{SO2} removal efficiency. As noted by Bolton (as cited by Liu \textit{et al.} \cite{liu_evaluation_2004}), the MSSS was developed to tackle the overprediction of the MPSS model. The present results are in line with those obtained by Liu \textit{et al.} \cite{liu_evaluation_2004} when using the MSSS in air. Namely, the MSSS model produces the lowest volume-average radiation intensity. In their work, the MSSS model was capable of reproducing experimental actinometric data better than the other models. Therefore, this model was chosen for the remainder parametric studies.

Overall, all simulations irrespective of the radiation model employed follow a similar trend. Initially, from molarity ratio 0 to 0.7, a linear response is observed. Increasing the molarity ratio leads to an increase in \ch{SO2} removal. For the models which predict a higher fluence rate the proportionality constant between molarity ratio and \ch{SO2} removal is higher and so the gradient of their trendline is steeper. From molarity ratio 0.7 to 1 the linearity decays progressively into a plateau reaching a maximum removal of 96-98\% depending on the model.

\paragraph*{Residence time.}

The efficiency of a UV reactor will be directly related to the ability of the photochemically active species to absorb UV radiation. Hence, increasing the exposure time of a particular species to UV light will improve photochemical conversion. This exposure can be evaluated by varying the residence time $\tau$ in seconds of the present simulations. Following the previous study of the different reactor models, the MSSS model was employed for this analysis. Other parameters and conditions remained the same.

\begin{figure}[h!]
\centering
        \includegraphics[width=0.7\textwidth]{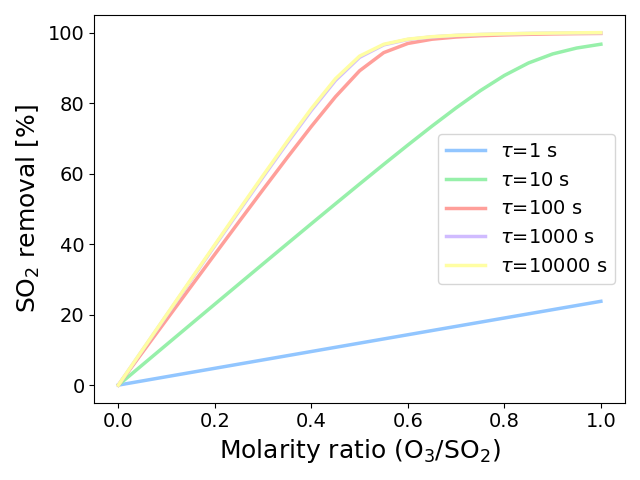}
        \caption{Evolution of the percentage removal of \ch{SO2} for different molarity ratios and different residence times $\tau$ in s. Note that the line corresponding to $\tau=1000$ s is barely visible as it lies just beneath the line corresponding to $\tau=10000$ s.}
        \label{fig:residence}
\end{figure}

Fig. \ref{fig:residence} shows the effect of the residence time on the overall removal efficiency of \ch{SO2}. As can be observed the residence time has a major impact on the resulting abatement of \ch{SO2}. For $\tau=1$ s just about 20\% removal is achieved with equal initial \ch{SO2} and \ch{O3} concentrations. Here, the slope is less than one. Statistically, this implies that for every 1 ppm of input \ch{O3}, less than one 1 ppm of \ch{SO2} is removed. Therefore, fast flow through the reactor is discouraged. Nevertheless, it should be noted that these idealized CSTR simulations assume ideal mixing. In reality, getting as close as possible to ideal mixing conditions will have to be done by turbulent mixing. At lower speeds, fluid flows tend to be laminar which do not promote mixing and so should be avoided. Hence, although higher residence times are ideal, these should not be achieved by lowering flow speeds. Therefore, a geometry-specific reactor has to be set up so that both relatively high residence times and turbulent mixing occur simultaneously. Increasing the residence time to $\tau=10$ s a notable increase in \ch{SO2} removal is achieved. At this time-scale, the slope of the graph approaches unity. Hence, the input concentration of \ch{O3} manages to remove close to that same concentration of \ch{SO2}. Therefore, close to 100\% removal is achieved at a molarity ratio equal to 1. Further increasing the residence time to $\tau=100$ s the value of the slope approaches a value of 2. This implies that any initial concentration of \ch{O3} manages to remove twice that concentration of \ch{SO2} from the gas phase. The peak close to 100\% removal is achieved in the vicinity of molarity ratio 0.5. This is the natural limit for this reaction, which can be seen by inspecting the original kinetic mechanism for the CSTR UV reactor in Table \ref{tab:mech}. As observed, under ideal conditions one mole of \ch{O3} produces one mole of oxygen singlets \ch{O(^1D)}. These singlets then react with water vapor to produce two moles of hydroxyl radicals \ch{^*OH} which finally react one-to-one with one mole of \ch{SO2}. This will happen if no other scavenging reaction occurs and the hydroxyl radicals only selectively oxidize \ch{SO2}. Under these conditions, one mole of \ch{O3} can oxidize two moles of \ch{SO2}. As seen in Fig. \ref{fig:residence}, this ideal limit can be approached with long residence times. However, increasing the residence time above $\tau=100$ s to $\tau=1000$ s or even $\tau=10000$ s produces diminishing returns. Therefore, the residence time for this reactor should be in the order of magnitude of $\tau=100$ s to achieve a good trade-off between high removal efficiency and reasonable residence times. 

\paragraph*{Reactor size.}

\begin{figure}[h!]
\centering
        \includegraphics[height=6.1cm]{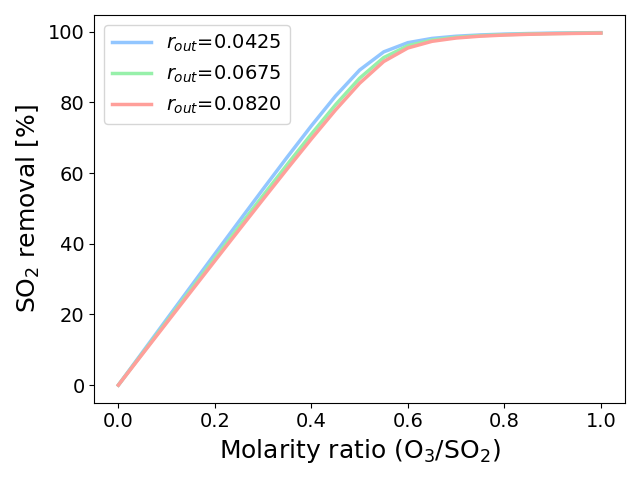}
        \includegraphics[height=6.1cm]{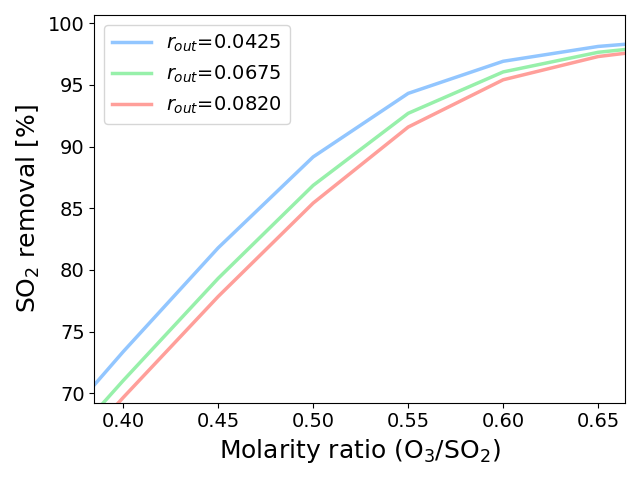}
        \caption{Evolution of the percentage removal of \ch{SO2} for different molarity ratios and different reactor sizes. \textit{Left}: Molarity ratio from 0 to 1. \textit{Right}: Zoomed view.}
        \label{fig:radius}
\end{figure}

Finally, following Montecchio \textit{et al.} \cite{montecchio_development_2018}, the size of the reactor was studied. This was done for three different outer radius sizes as shown in Table. \ref{tab:params}. Fig. \ref{fig:radius} shows the results of this parameter study. As can be observed, the impact of the reactor size for the selected dimension is not very large for molarity ratios below 0.2 or above 0.8. Around molarity ratio equals 0.5 the effect of the reactor size is more visible. Here, the smaller reactor is able to perform better and remove a higher percentage of initial \ch{SO2}. Increasing the reactor outer radius will in turn increase the reactor volume and decrease the abatement efficiency. Therefore, to improve the removal of \ch{SO2}, the aspect ratio of the reactor must be large, with the axial length being much larger than the radial length.

\subsubsection*{3.4. Time-dependent results}
\addcontentsline{toc}{subsubsection}{3.4. Time-dependent results}

To conclude the results section, some detailed results with temporal resolution will be shown. Unlike previous sections, results shown here are one for a single simulation with the optimum or desired conditions obtained from the parametric studies. This includes the MSSS model, the smallest reactor size, and a residence time of $\tau=100$ s. Here, a molarity ratio of 0.5 was chosen. Results with temporal resolution enable an understanding of the time scales of different reactions.

\begin{figure}[h!]
\centering
        \includegraphics[height=6.1cm]{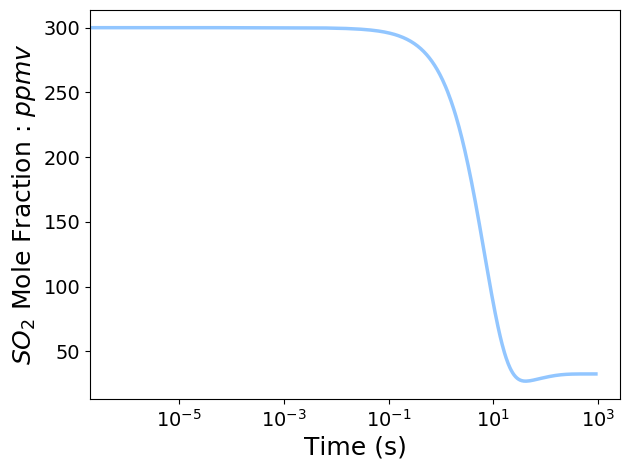}
        \includegraphics[height=6.1cm]{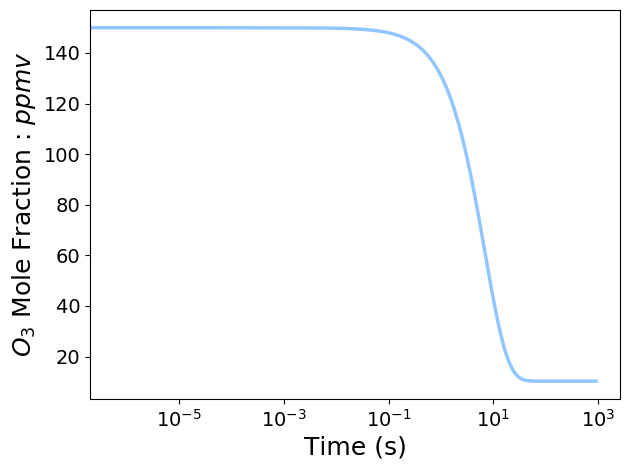} \\
        \includegraphics[height=6.1cm]{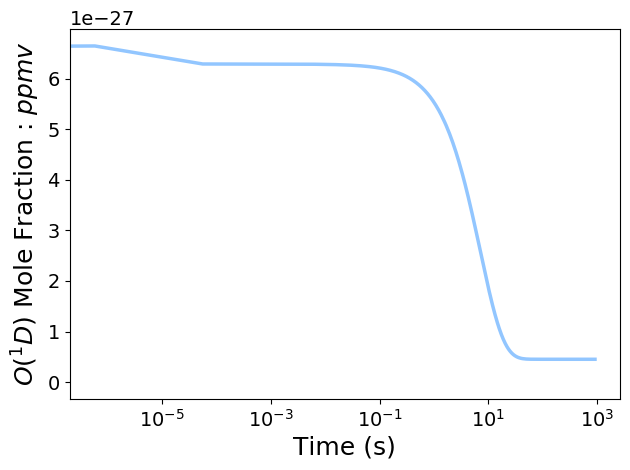}
        \includegraphics[height=6.1cm]{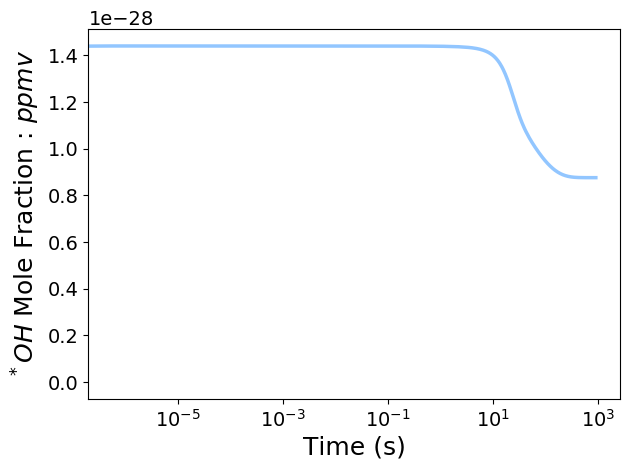} \\
        \includegraphics[height=6.1cm]{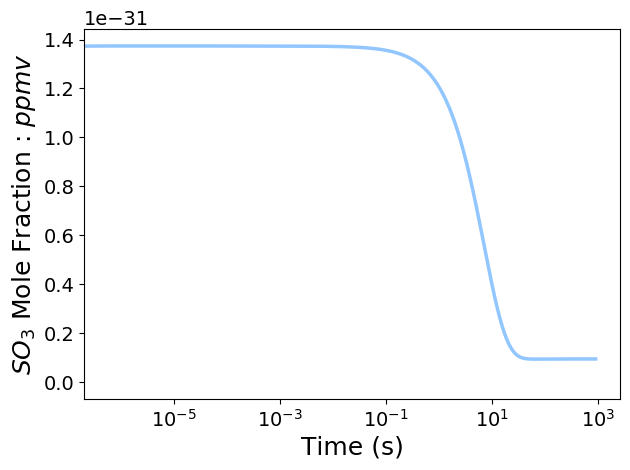}
        \includegraphics[height=6.1cm]{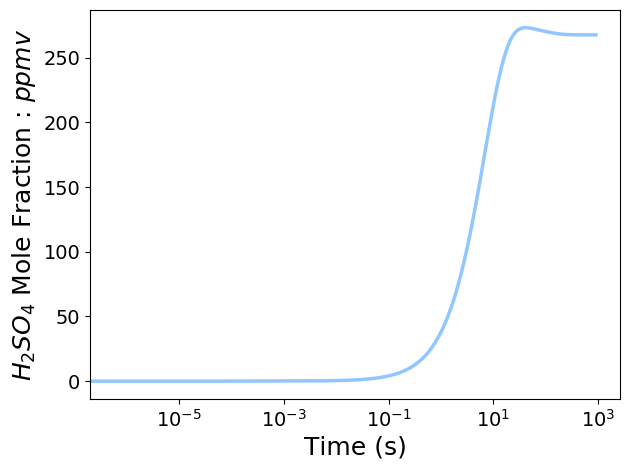} \\        
        \caption{Temporal evolution of the concentration of different important species for the CSTR UV simulation.}
        \label{fig:species}
\end{figure}

Fig. \ref{fig:species} shows the time dependant concentration of six relevant species in the UV reactor. First, \ch{SO2} and \ch{O3} can be seen to follow a similar trend as one is being converted to oxygen singlets \ch{O(^1D)} by the UV light while the other is being oxidized by the hydroxyl radicals \ch{^.OH}. The time-scale at which both take place appears to be almost identical, which implies that the intermediate steps that are between \ch{O3} depletion and \ch{SO2} removal happen fast. This hypothesis is further substantiated by observing the evolution of \ch{O(^1D)} and \ch{^.OH}. As can be seen, both remain at negligible concentration values which implies that all molecules being generated are instantaneously being consumed. Something similar happens to \ch{SO3} which is an intermediate species. After \ch{SO2} reacts with \ch{^.OH} it produces \ch{HSO3} which reacts with the abundant \ch{O2} to produce \ch{SO3}. However, again this species is being consumed at least as fast as it is being produced due to its constant low concentration. When \ch{SO3} reacts with the water vapor present in the gas mixture it is converted into \ch{H2SO4} which can be also seen in Fig. \ref{fig:species} in much higher concentrations. 

\subsection*{4. Conclusions and future outlook}
\addcontentsline{toc}{subsection}{4. Conclusions and future outlook}

In the current study, simulations of a CSTR reactor were performed to analyze the abatement of \ch{SO2} by \ch{O3} via UV light. To do so, an annular reactor was discretized and several radiation models were employed to simulate the volume-average radiation intensity. The theory employed has been described in detail and equations have been derived. The information regarding radiation distribution was employed to numerically evaluate the reaction rate of a photochemically active step. This step converted \ch{O3} to oxygen singlets which gave rise to hydroxyl radicals which end up oxidizing the main pollutant \ch{SO2}. The fundamental concepts of photochemical modeling were presented. The chemical process was simulated using the Cantera \cite{goodwin_cantera_2016} software in python \cite{van_rossum_python_2009}. For this, a chemical kinetic mechanism was required, which was adapted from Meusinger \textit{et al.} \cite{meusinger_treatment_2017}.

To understand and minimize the error committed in the simulations, a grid convergence study was undertaken. A low number of mesh elements ($\sim 200$) were found to be enough for obtaining low errors. Some radiation models require the discretization of the UV lamp into point light sources or segments. To obtain acceptable errors and minimize the computational cost, a method to obtain the optimum value of the discrete light sources to be used was developed. It was shown that for the present configuration, a low number of light sources ($< 50$) was enough to obtain errors smaller than 1\%.

The results presented here include a parametric study on several simulation variables. When the different models were assessed the LSI and MPSS models were observed to predict similarly high overall \ch{SO2} removal. Following, the RAD and RAD-LSI models presented lower removal due to their mainly radial conceptualization of the direction of the emitted UV light. Finally, the MSSS method produced the lowest removal rate of \ch{SO2} as was expected from the literature. The residence time was also studied. In this parametric analysis, it was observed to be the most impact variable. Low residence times ($1 < \tau < 10$ s) yielded poor pollutant removal rates. Very high residence times ($1000 < \tau < 10000$ s) produced much higher \ch{SO2} removal but results were almost identical, showing diminishing returns for such large values. Hence a moderate residence time of $\tau=100$ s was observed to be a reasonable compromise. The reactor size was also studied. This variable was seen to have a moderate effect on removal rates for the range studied. Nevertheless, it was concluding that reactors that have a low radial length compared to their axial lengths will perform better. Hence, large aspect ratios are desirable. Finally, time-dependent results were presented. Here, the evolution of some of the key species in the present chemical process was studied. It was shown that some intermediate species such as \ch{O^1D}, \ch{^.OH} and \ch{SO3} are consumed at an equal or faster rate than they are produced. All in all, results show that the application of the present methods constitutes a viable strategy for gas-phase pollutant removal systems. 

Further developments to the present work include several aspects not directly evaluated here. For instance, a validation case could be set up. Chemical UV reactors are dependant on many aspects such as geometry, chemical species, UV lamp characteristics, and placement. Hence, either another detailed experimental results should be replicated or a new experimental work should be undertaken. A more complex version of the chemistry employed here could be studied. Hence, more complex mechanisms should be studied to assess how performance varies with an increasing number of intermediate species and/or scavenging molecules. Similarly, this reactor could be configured in series with a \ch{NO_x}-\ch{O3} system. This will both simulate a more realistic scenario - as \ch{NO_x} and \ch{SO_x} tend to coexist in exhaust gases - and evaluate the selectivity of hydroxyl radical oxidation. Here, the reflection of UV light from the reactor walls has not been addressed. This process would require experimental measurements of the material, roughness, and fouling over time of the wall. Furthermore, the effect of several UV lamps and their placing could be evaluated. Finally, scrapping the idealization of chemical homogeneity from perfect mixing could lead to more realistic simulations. This would entail more complex and time-consuming computational fluid dynamics simulations (CFD) coupled with both the local radiation model and the local chemistry kinetic rates. Under this framework, more detailed radiation models could be studied (i.e. the discrete ordinate method \cite{liu_evaluation_2004}) and more complex turbulent modeling such as large-eddy simulations (LES) could be employed. 

\subsection*{Acknowledgements}
\addcontentsline{toc}{subsection}{Acknowledgements}

This work has been funded by ÅForsk (grant 19-422) with complementary financing from Formas (Swedish Research Council for Sustainable Development).

\printbibliography

\end{document}